\documentclass{JINST}
\usepackage{lineno}
\usepackage[percent]{overpic}

\newcommand{\DBD}{0$\nu$DBD}
\newcommand{\PO}{$^{210}$Po}

\newcommand{\TEO}{$\mathrm{TeO}_2$}
\newcommand{\TEHT}{$^{130}\mathrm{Te}$}

\newcommand{\THO}{$^{232}\mathrm{Th}$}

\newcommand{\Cuore}{CUORE}

\providecommand*{\un}[1]{\ensuremath{\mathrm{\,#1}}}

\title{Lowering the energy threshold of large-mass bolometric detectors}

\author{S. Di Domizio$^a$, F. Orio$^b$~ and M. Vignati$^b$\thanks{Corresponding
author.}\\
\llap{$^a$}Universit\`a di Genova and Sezione INFN di Genova, Genova
  I-16146, Italy\\
\llap{$^b$}Sapienza Universit\`a di Roma and Sezione INFN di Roma, Roma
  I-00185, Italy\\
  E-mail:\email{marco.vignati@roma1.infn.it}
}

\abstract{Large-mass bolometers are used in particle physics experiments to
search for rare processes. The energy threshold of such detectors plays a critical role in
their capability to search for dark matter interactions and rare nuclear decays. We have developed a
trigger and a pulse shape algorithm based on the matched filter technique which, when  
applied to data from test bolometers of the \Cuore\ experiment, lowered the
energy threshold from tens of keV to the few keV region. The detection
efficiency is in excess of 80\%, and nearly all nonphysical pulses are rejected.}


\begin{document}

\section{Introduction}

Bolometers are detectors in which the energy from particle interactions is converted
into heat and measured via the resulting rise in temperature. They provide excellent
energy resolution, though their response is slow compared to conventional detectors.
These features make them a suitable choice for experiments searching for rare
processes, such as neutrinoless double beta decay (\DBD) and dark matter (DM)
interactions.

The \Cuore\ experiment will search for \DBD\ of
\TEHT~\cite{Ardito:2005ar} using an array of 988 \TEO\ bolometers
of 750\un{g} each. Operated at a temperature of about 10\un{mK}, these
detectors maintain an energy resolution of a few keV over their energy
range, extending from a few keV up to several MeV.  The measured resolution
in the region of interest (2527\un{keV}) is about 5\un{keV\,FWHM}; this,
together with the low background and the high mass of the experiment,
determines the sensitivity to the \DBD.  \Cuore\ could also search for DM
interactions, provided that the energy threshold is sufficiently low. DM
candidates such as weak interacting massive particles (WIMPs) and axion-like particles (ALPs)~\cite{Bertone:2004pz} are expected to
produce signals at energies below $\sim 30\un{keV}$, and the interaction
rate increases as the energy decreases.

The existing \Cuore\ energy threshold, achieved using a trigger algorithm
applied to the raw data samples, is of the order of tens of keV. 
If the threshold were of few keV, 
the \Cuore\ experiment could be sensitive to DM interactions
and thus play an important role in this growing research area.  At low
energies the detection capability is limited by the detector noise.
Electronics spikes, mechanical vibrations, and temperature fluctuations can
produce pulses that, if not properly identified, generate nonphysical
background. The lower the energy released in the bolometer, the
more difficult it is to discriminate between physical and nonphysical pulses.

In this paper we present a method to lower significantly the energy threshold, and a pulse shape parameter featuring high rejection power even at low energy.
The advantage of this method comes from running the trigger and pulse shape algorithms on data that have been previously processed with the matched filter technique~\cite{Gatti:1986cw}.
A lower energy threshold can be obtained because, compared to raw data, matched filtered data feature a higher signal to noise ratio.
The algorithms need as input the expected shape of the signal and the noise power spectrum of the bolometer.
No manual tuning is needed, because all the parameters can be set at optimal
values automatically.
Application of this technique to data from a test detector shows that the energy threshold can be lowered to a few keV, and that nonphysical pulses can be efficiently identified.
The technique we have developed can be used with any kind of bolometric detector.

\section{Experimental setup} 

The algorithms developed in this paper were tested on two
bolometers operated by the \Cuore\ collaboration at the LNGS underground
laboratory. The main purpose of the test was to check one of the first 
production batches of \Cuore\ crystals~\cite{ioanprod}.  
In the following we will describe our studies taking the first bolometer as 
example, while the final results will be shown for both.

A \Cuore\ bolometer is composed of two main parts,
a \TEO\ crystal and a neutron transmutation doped Germanium (NTD-Ge) thermistor~\cite{wang,Itoh}. The crystal is
cube-shaped (5x5x5\un{cm^3}) and held by Teflon supports in copper frames.  The frames
are connected to the mixing chamber of a dilution refrigerator, which
keeps the system at the temperature of $\sim 10\un{mK}$.  The thermistor
is glued to the crystal and acts as thermometer. When energy is released
in the crystal, its temperature increases and changes the thermistor's
resistance~\cite{Mott:1969,efros,Itoh:1996}.  The thermistor is
biased with a constant current, and the voltage across it constitutes the
signal~\cite{AProgFE}.  Usually, a Joule heater is also glued to the crystal.
It is used to inject controlled amounts of energy into the crystal, to emulate
signals produced by particles~\cite{stabilization,Arnaboldi:2003yp}.

The typical response of \Cuore\ bolometers to particles impinging on the crystal is of order $100\un{\mu V/MeV}$. The signal frequency bandwidth is
$0-20\un{Hz}$, while the noise components extend to higher frequencies.
The signal is amplified, filtered with a 6-pole active Bessel filter with
a cut-off frequency of 12\un{Hz} and 20\un{Hz} for the first and the second bolometer, respectively,
 and then acquired with an 18-bit ADC
with a sampling frequency of 125\un{Hz}.
Given the low event rate and the small frequency bandwidth of the
signal, it is possible to save the complete continuous data stream
of the detector voltages, so that it can be reprocessed offline.  An online
software trigger is used as well: when it fires, a fixed number of data
samples is saved.  There are three kinds of triggers: those caused by
particle-like pulses, random waveform samplings, and those generated by
the Joule heater.  Each event is tagged with a flag that indicates
the trigger type.  The standard online particle
trigger algorithm is based on a simple derivative approach: it fires
when the slope of the signal remains above threshold for a certain amount of time.
In the measurements discussed in this paper, the energy threshold reached with this algorithm was 11\un{keV} on the first bolometer and 40\un{keV} on the second one.

\begin{figure}[tbp]
\centering
\includegraphics[width=0.65\textwidth]{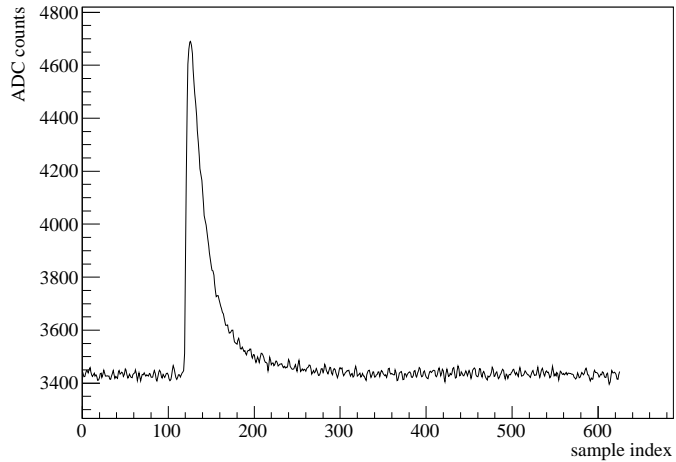}
\caption{Signal produced by a 88\un{keV} $\gamma$
particle fully absorbed in the detector.  The signal is sampled at
125\un{Hz} and the length of the window corresponds to 5.008\un{s} (626 samples).
\label{fig:signal_88} 
}
\end{figure}

\begin{figure}[tb] \centering
\includegraphics[width=0.65\textwidth]{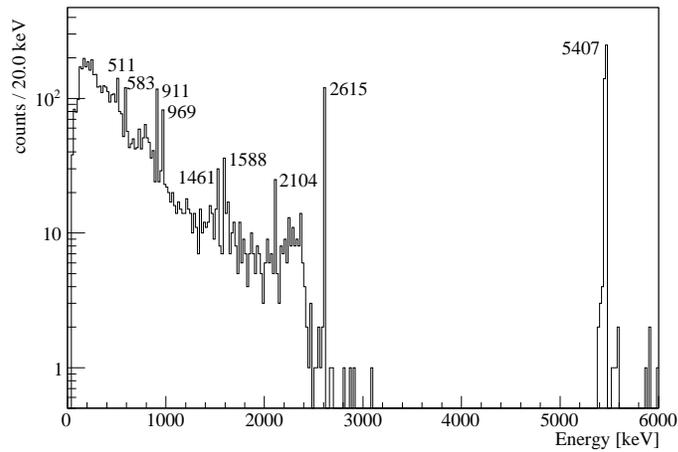}
\label{fig:calspectrum} 
\caption{Energy spectrum of the calibration measurement. All lines are generated
by the \THO\ calibration source except for the lines at 1461\un{keV}, due
to $^{40}$K contamination of the refrigerator, and at 5407\un{keV}, due
to $^{210}$Po contamination in the \TEO\ crystal.}
\end{figure}

A typical signal produced by an 88\un{keV} $\gamma$ particle
generated by a $^{127m}$Te de-excitation in the crystal is shown in
Fig.~\ref{fig:signal_88}.  The length of the acquisition window was
5.008\un{s}, corresponding to $M=626$ samples.
The energy calibration was made by exposing the bolometer to  a \THO\
source; for what concerns our studies, the calibration function
can be considered linear at energies below 100\un{keV}.  For the first
bolometer the conversion factor from ADC~counts to energy is $\sim 15
\un{ADC\,counts / keV}$.
The acquired calibration spectrum is shown in Fig.~\ref{fig:calspectrum}.
In it, the most prominent peaks of the \THO\ source are visible, together with two
lines at 1461\un{keV} and 5407\un{keV},
which are due to $^{40}$K contamination of the refrigerator
and $^{210}$Po contamination of the crystal
bulk, respectively. The $^{210}$Po line decays with a half-life of 138\un{days} and is visible
in recently-produced crystals. In the data analyzed in this paper, the  rate of the $^{210}$Po
line was about 10\un{mHz} and significantly affected the detection
efficiencies at low energies, as will be shown in the next sections.

\section{Data filtering}

\begin{figure}[htbp]
\centering
\begin{minipage}{0.497\textwidth}
\includegraphics[clip=true,width=1\textwidth]{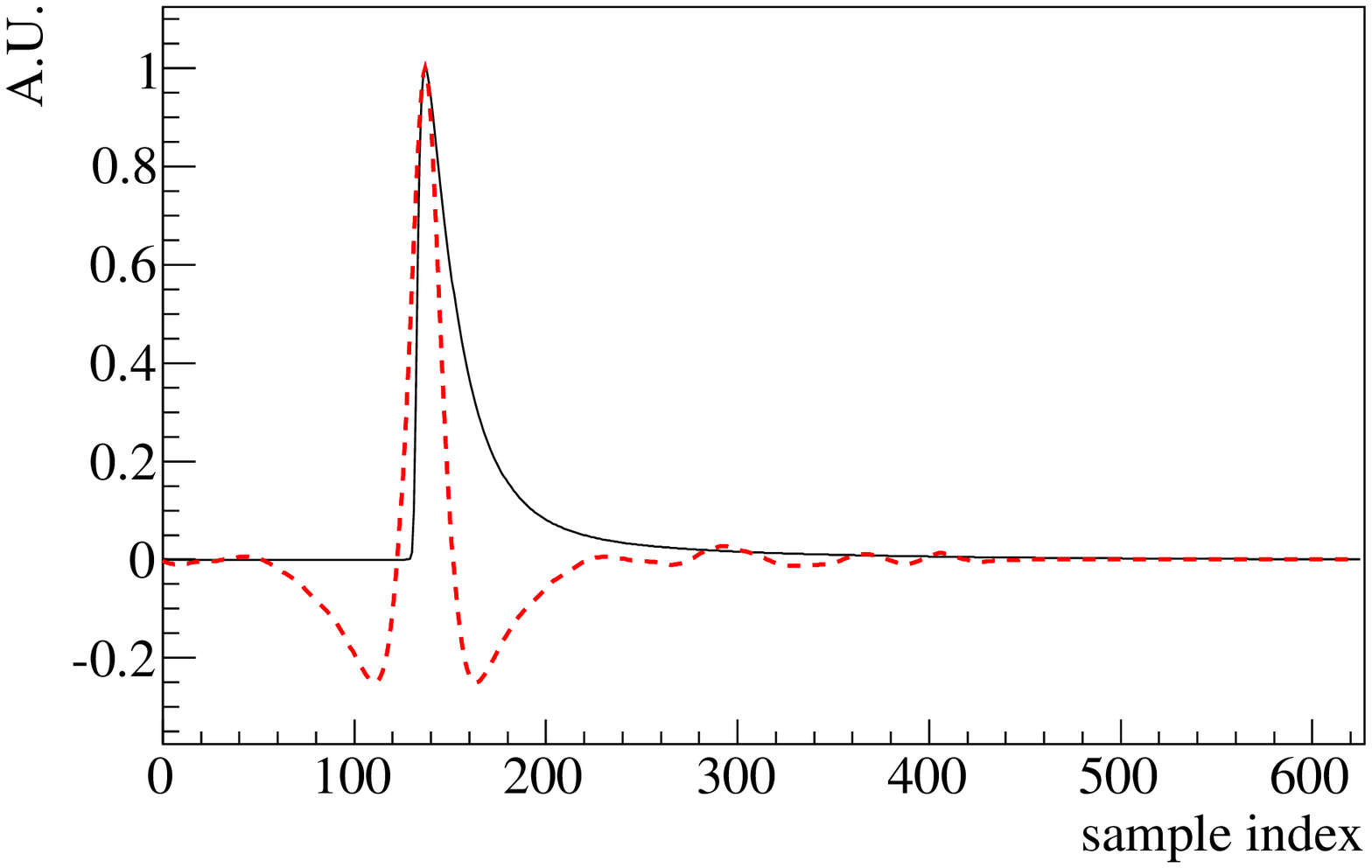}
\end{minipage}
\begin{minipage}{0.497\textwidth}
\includegraphics[clip=true,width=1\textwidth]{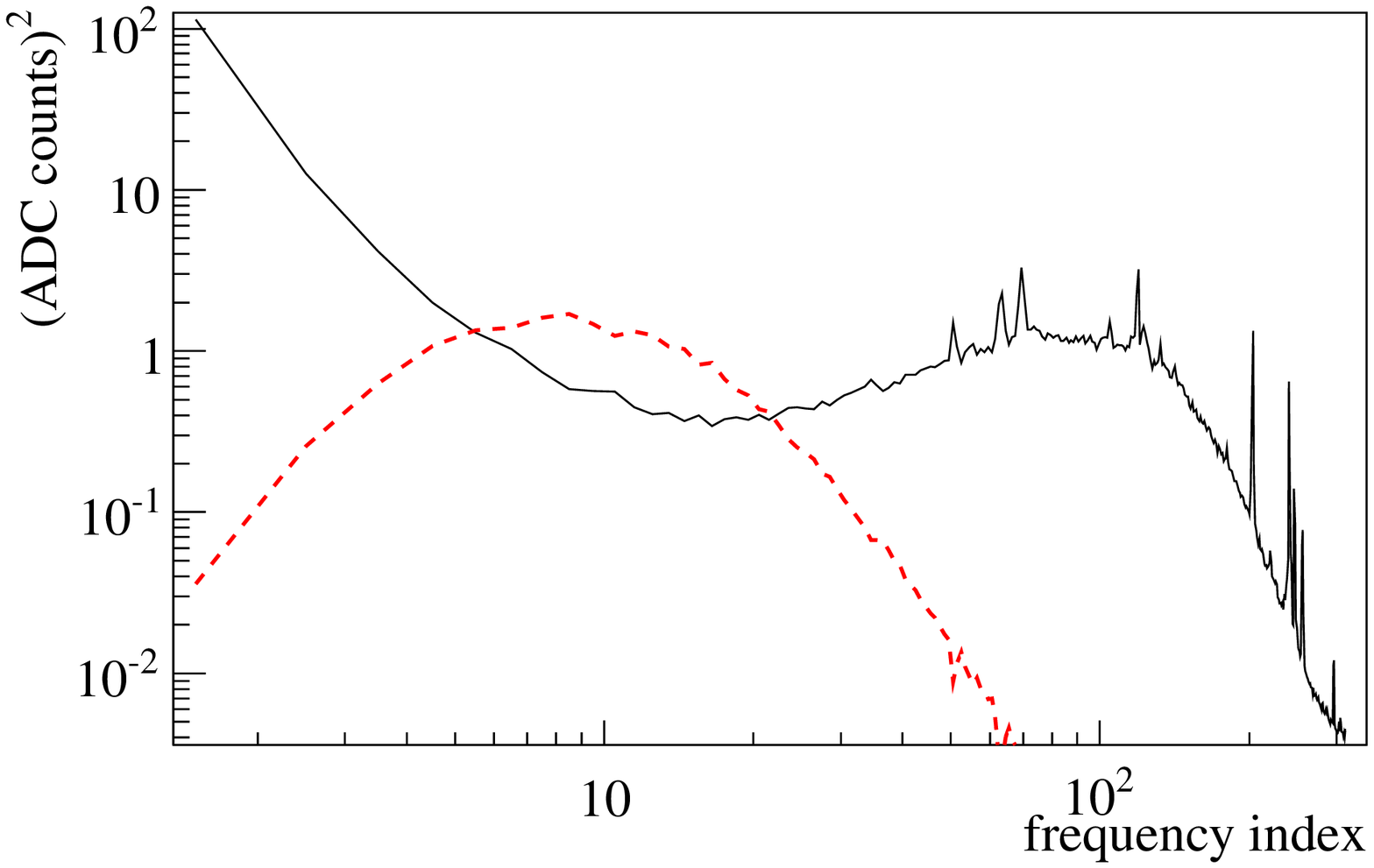}
\end{minipage}
\caption{
Average pulse (left) and noise power spectrum (right) of the detector, before (solid black line) and after (dashed red line) the filter. The filtered average pulse is symmetric and has the same amplitude as the original. 
The filtered noise is enhanced in the signal frequency bandwidth and suppressed elsewhere.
}
\label{fig:apnps}
\end{figure}

The trigger and the pulse shape identification algorithms presented in
this paper operate on data samples processed using the matched filter
technique~\cite{Gatti:1986cw}. This filter is designed to estimate the amplitude of a signal,
maximizing the signal to noise ratio. The transfer function is matched
to the signal shape, and pulses with different shape are suppressed.
The noise power spectrum of the detector, $N(\omega_k)$, and the signal
shape, $s_i$, are needed to build the transfer function:
\begin{equation}
H(\omega_k) =  h \frac{s^*(\omega_k)}{N(\omega_k)}e^{-\jmath\, \omega_k i_M }\,,
\label{eq:of}
\end{equation}
where $s(\omega_k)$ is the Discrete Fourier Transform (DFT) of $s_i$, $i_M$ is the maximum position of $s_i$
in the acquisition window, and $h$ is a normalization constant that 
leaves unmodified the amplitude of the signal:
\begin{equation}
h = 1 / \sum_k \frac{|s(\omega_k)|^2}{N(\omega_k)}\,.
\end{equation}
The noise power spectrum at the filter output can be estimated as
\begin{equation}
N_f(\omega_k) = h^2   \frac{|s(\omega_k)|^2}{N(\omega_k)}\,,
\end{equation}
from which the resolution $\sigma_f$ can be derived:
\begin{equation}
\sigma_f^2 = \sum_k N_f(\omega_k) = h\,.
\label{eq:of_noise_ps}
\end{equation}
The estimation of $N(\omega_k)$ is made by averaging the power spectrum of a large set of data windows not containing signals, while the signal shape $s_i$ is obtained by averaging a set of particle signals selected from the calibration measurement in the 1-3\un{MeV} range.
The filtered average pulse is symmetric and has the same amplitude as the
original (see Fig.~\ref{fig:apnps}, left). On the other hand, noise 
frequencies with low signal to noise ratio are suppressed (see Fig.~\ref{fig:apnps}, right).
The resolution before the filter, $\sigma$, is $1.1\un{keV}$ and,  according to Eq.~\ref{eq:of_noise_ps}, is reduced by the filter to  $\sigma_f = 0.3\un{keV}$.

The filtering of the data samples can be implemented in two ways,
in the time domain or in the frequency domain. In the time domain
one has to implement a finite impulse response (FIR) filter whose kernel is the DFT of Eq.~\ref{eq:of}.
In the frequency domain, data samples first have to be transformed using DFT,
then multiplied by Eq.~\ref{eq:of}, and then transformed back in the time domain. 
The number of floating-point operations of the two methods is of the same order of magnitude, namely $O(M^2)$.
However, in the frequency domain convolution, FFT algorithms can be used in place of the standard DFT, reducing the number
of operations from $O(M^2)$ to $O(M\log M)$. In view of a large scale application, we chose to filter
the data in the frequency domain.

DFT algorithms assume that the window to be filtered is periodic.  
If the input data are not periodic, the convolution
will falsely pollute the left side of the window with data from the
right side and vice versa.  This problem is known as ``DFT convolution
wraparound''~\cite{nr} and can be solved by padding the response function of the filter
with zeros, such that a slice of the window is correctly filtered.
We follow these steps to build the transfer function:
\begin{enumerate}
\item Compute the matched filter transfer function $H(\omega_k)$ of length
$M$ using the average signal shape $s_i$ and the noise power spectrum $N(\omega_k)$,
both of length $M$.
\item Transform $H(\omega_k)$ into the time domain, obtaining
the response function $H_i$ of length $M$.
\item Insert $M$ zeros in the middle of $H_i$, obtaining the response function
$H^d_{i}$ of length $2M$.
\item Smooth $H^d_i$ in proximity of the zero insertion,
obtaining the smoothed response function $H^{ds}_i$. 
\item Transform $H^{ds}_{i}$ into the frequency domain, obtaining the
final transfer function $H^{ds}(\omega_k)$, of length $2M$.
\end{enumerate}
The smoothing of $H^d_i$ is needed to avoid the Gibbs
phenomenon~\cite{gibbs1898fourier}: if $H^d_i$ does not approach zero 
where the zeros will be inserted, a discontinuity will be created,
introducing fake oscillations in its DFT. 
The smoothing is made by multiplying $H^d_i$ by a quarter-period cosine shape.
The cosine period is $L=120$ samples, and the quarter-period that goes from one to zero is used to smooth
$H^d_i$ at the left of the zero insertion, while the  quarter-period that goes from zero to one is used to smooth $H^d_i$ at the
right of the zero insertion.

\begin{figure}[htb]
\centering
\includegraphics[width=0.45\textwidth]{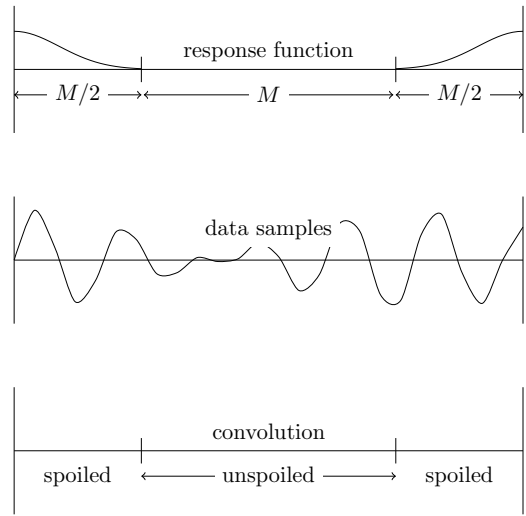}
\caption{Padding of the filter response function. To avoid the DFT wraparound problem
the original response function is padded in the middle with a number of zeros equal
to its length. In this way the product of the response function DFT and
of the data samples DFT produces a correct convolution in correspondence
of the response function zeros.} \label{fig:of_noiwrap}
\end{figure}

\begin{figure}[bt]
\centering
\includegraphics[width=0.5\textwidth]{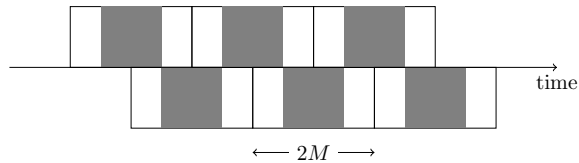}
\caption{Filtering of the continuous data flow. Each window has size $2M$
(white rectangle), and the $M$ middle samples are correctly filtered
(gray rectangle). The windows are overlapped by $M$, such that the $M$
middle samples of each window are concatenated.} \label{fig:of_overlapped}
\end{figure}

When filtering a data sample of length $2M$, the first and
the last $M/2$ samples will be spoiled while the $M$ middle  samples
will be correctly filtered (Fig.~\ref{fig:of_noiwrap}).
The continuous data flow is processed by filtering windows of length $2M$
overlapped by $M$ and taking the $M$ middle samples of each window as
output. The concatenation of the middle samples produces
a continuous flow of correctly filtered data (Fig.~\ref{fig:of_overlapped}).

\section{Trigger algorithm}
\begin{figure}[htbp]
\centering
\begin{minipage}{0.49\textwidth}
\begin{overpic}[width=1\textwidth]{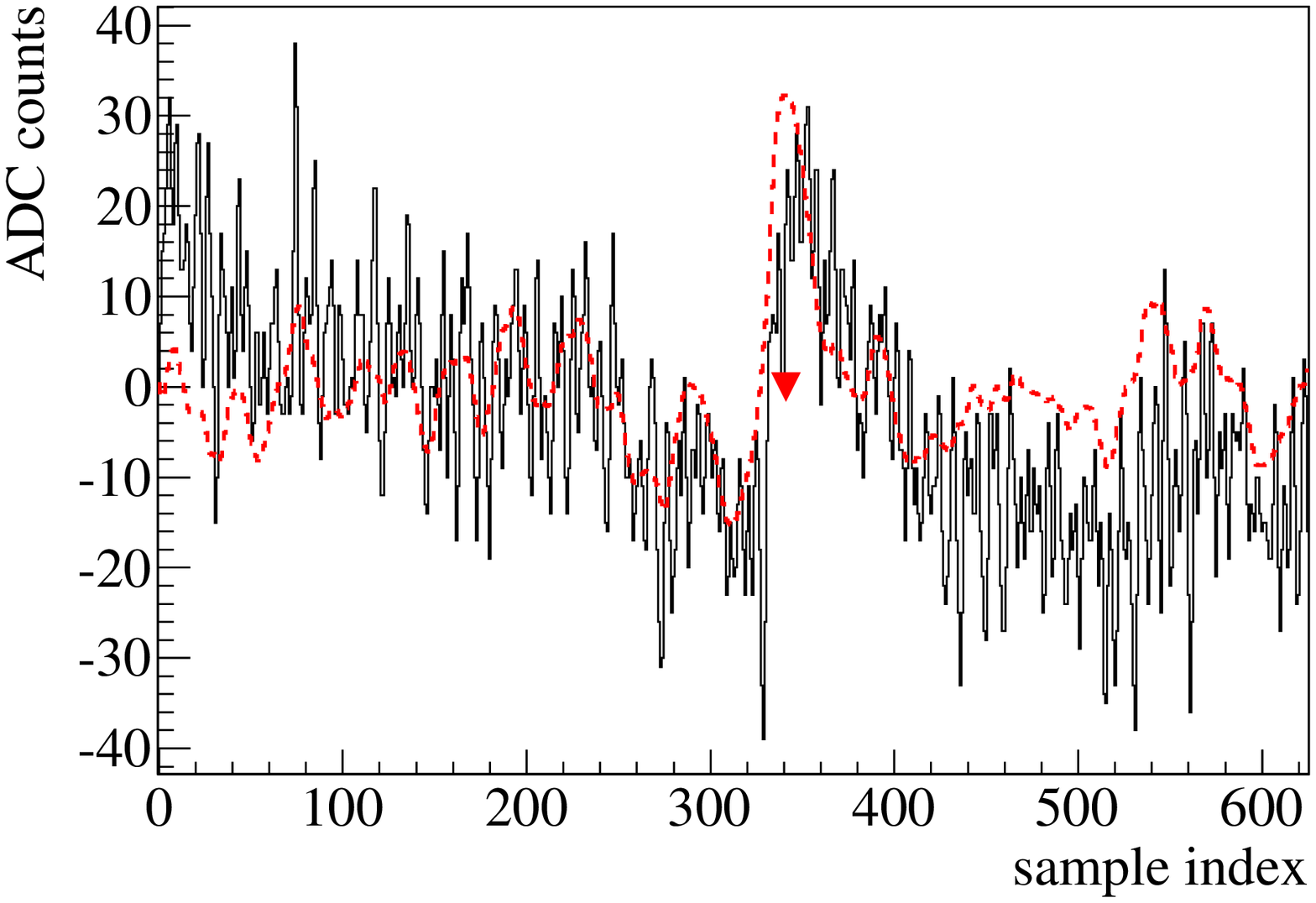}
\put(60,60){\footnotesize 3\un{keV} signal}
\end{overpic}
\end{minipage}
\begin{minipage}{0.49\textwidth}
\begin{overpic}[width=1\textwidth]{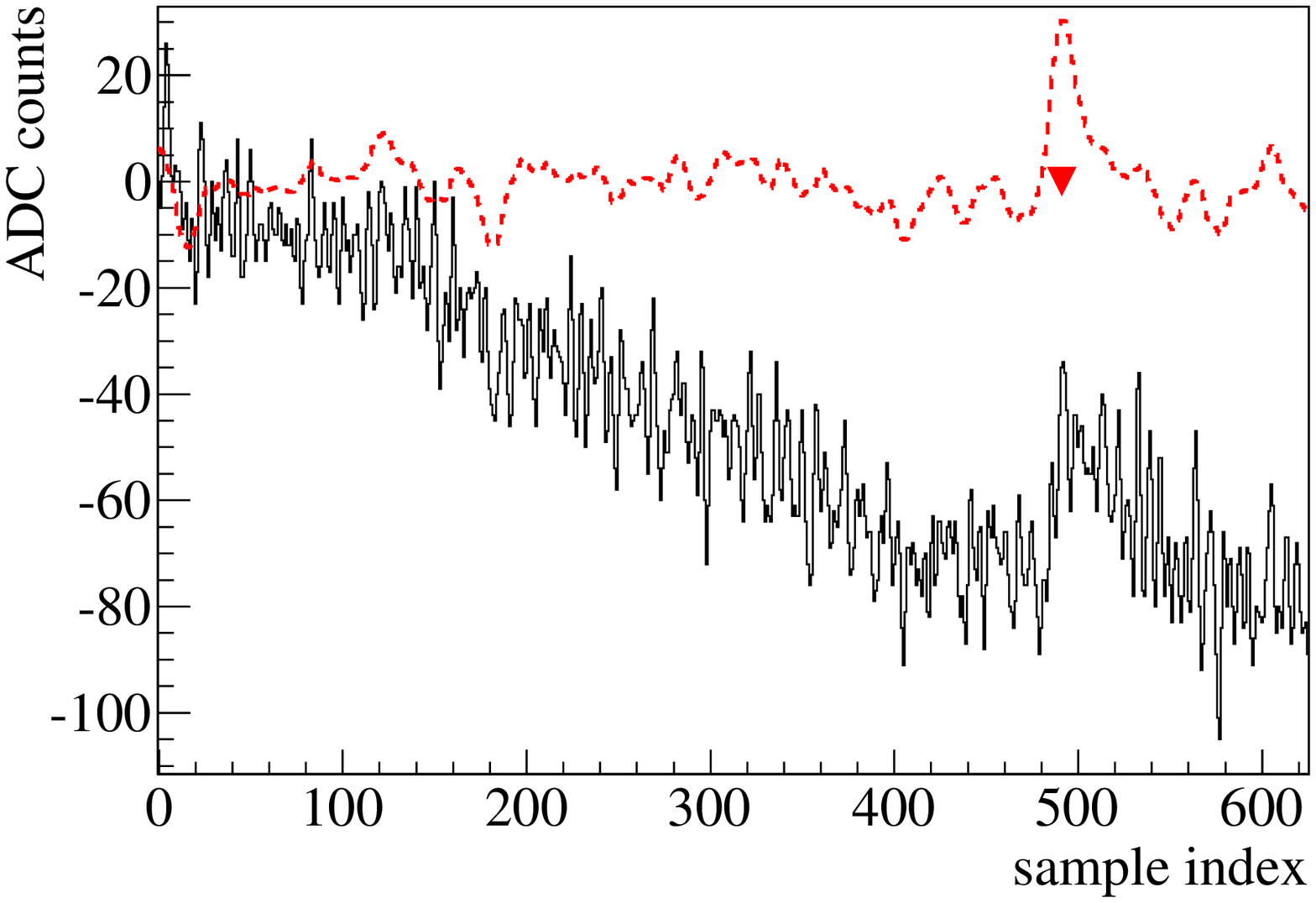}
\put(40,60){\footnotesize 3\un{keV} signal}
\end{overpic}
\end{minipage}
\begin{minipage}{0.49\textwidth}
\begin{overpic}[width=1\textwidth]{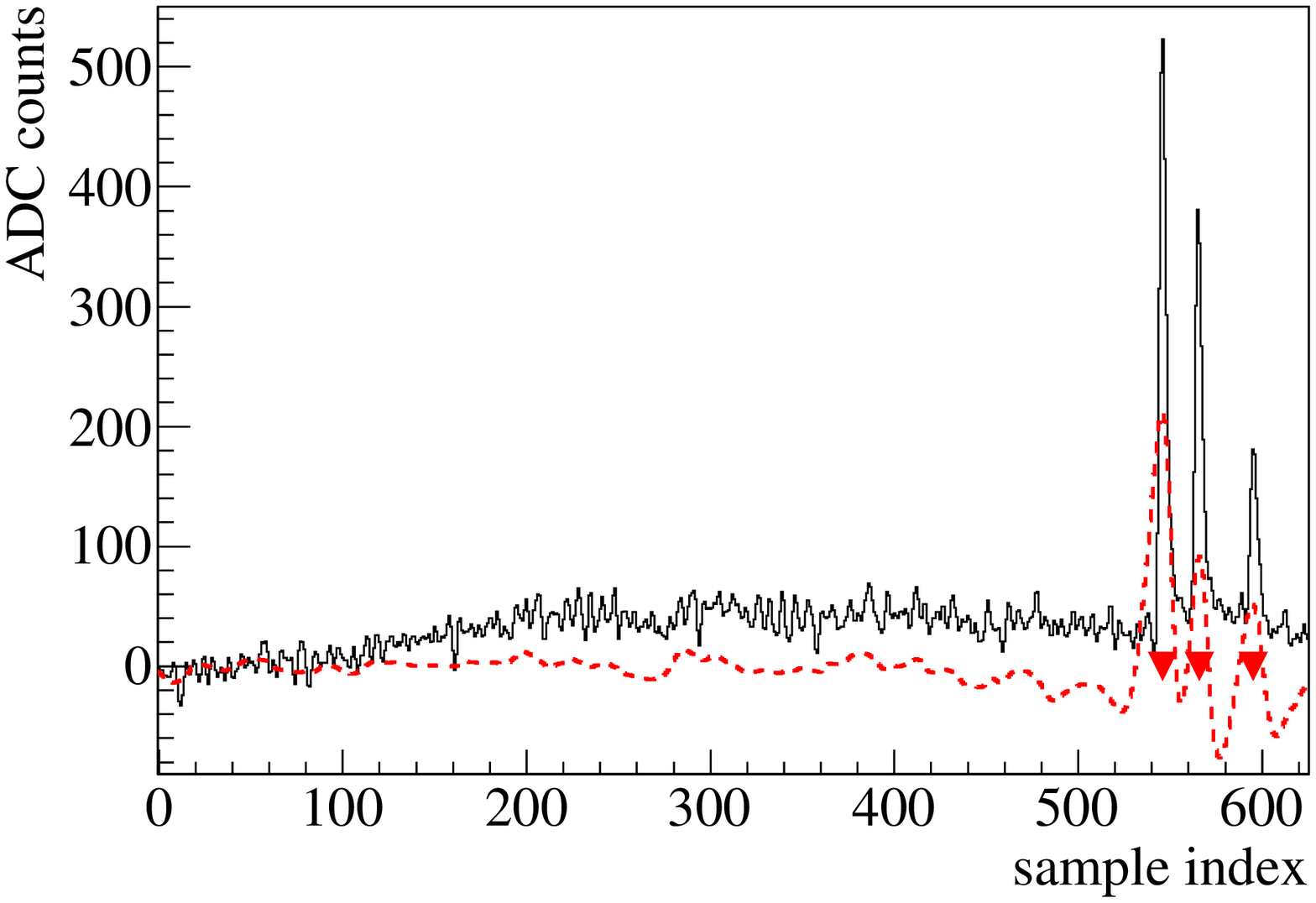}
\put(40,60){\footnotesize noise spikes}
\end{overpic}
\end{minipage}
\begin{minipage}{0.49\textwidth}
\begin{overpic}[width=1\textwidth]{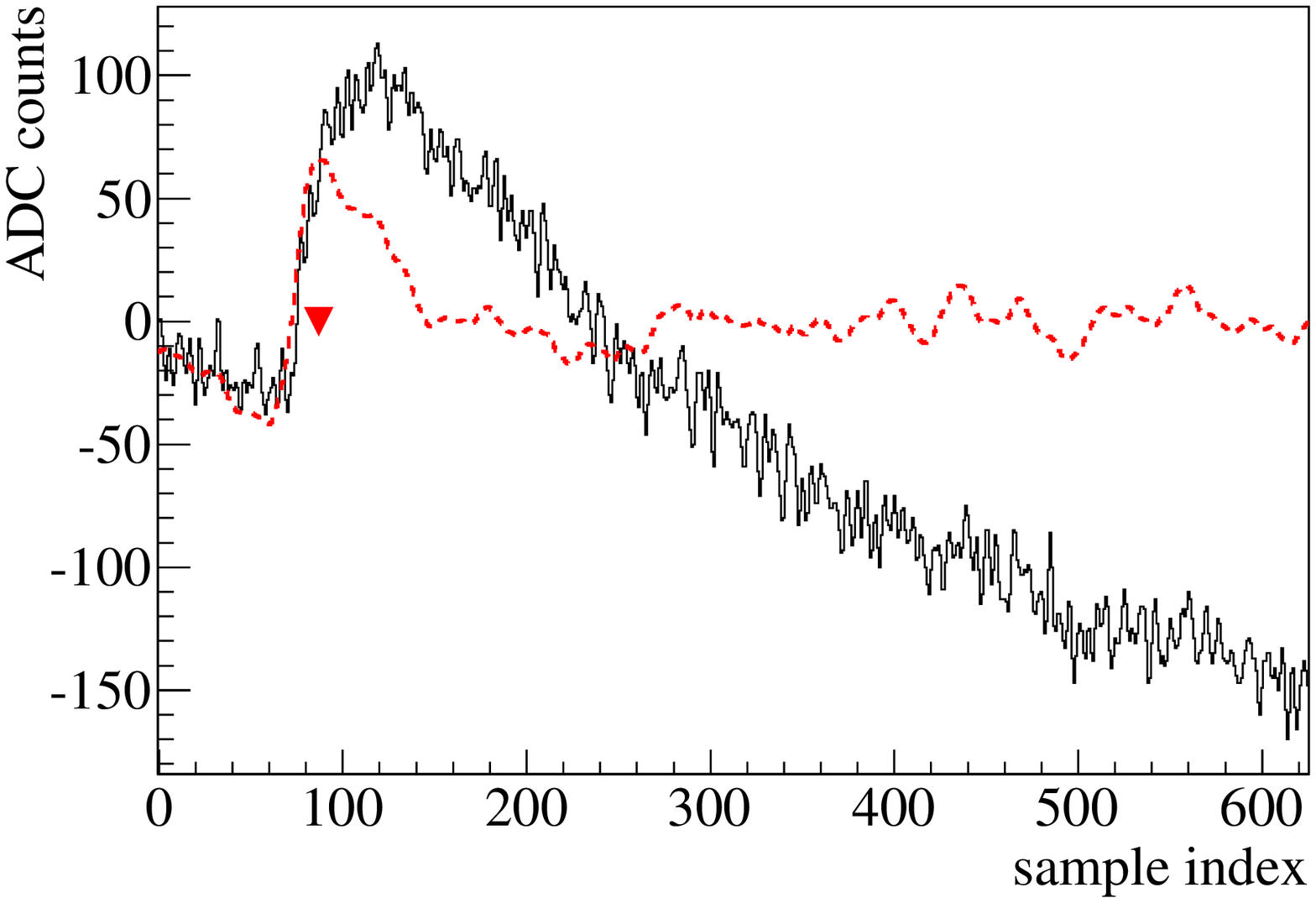}
\put(40,60){\footnotesize detector vibration}
\end{overpic}
\end{minipage}
\caption{Windows of data samples containing pulses. The raw data (solid black line) have been shifted such that the first sample has zero ADC counts; the filtered data (dashed red line) are not modified.
Red triangles identify the pulses detected by the trigger algorithm.
The filter removes the baseline drifts and suppresses pulses with different shapes than the expected signal.
}
\label{fig:ot_filt_windows}
\end{figure}

Some slices of the filtered data are shown in Fig.~\ref{fig:ot_filt_windows} and
exhibit the following features:
\begin{enumerate}
\item The noise fluctuations are reduced.
\item The baseline is zero.
\item The filter is sensitive to the shape of the expected signal and 
      pulses with different shape are suppressed.
\end{enumerate}

Pulses are triggered when the data samples exceed a positive
threshold, and a new trigger is possible when the data samples return
below threshold, or after that a local maximum is found. 
The trigger threshold $\theta$ is defined in
terms of number of sigma of the filtered noise ($\sigma_f$), which is known a priori
(see Eq.~\ref{eq:of_noise_ps}). The triangles
in Fig.~\ref{fig:ot_filt_windows} show the found triggers and their
positions, using the threshold $\theta = 4.5\sigma_f$.
The trigger position is then shifted with respect to the maximum, 
so that it lies in the middle of the rise of the pulse. This operation is needed
since in the data analysis the trigger is expected to lie before
the maximum of a signal. The size of the shift back is fixed and 
is computed from the time difference between the middle of the rise
and the maximum of the non-filtered average pulse. 

\begin{figure}[t]
\centering
\begin{minipage}{0.49\textwidth}
\includegraphics[clip=true,width=1\textwidth]{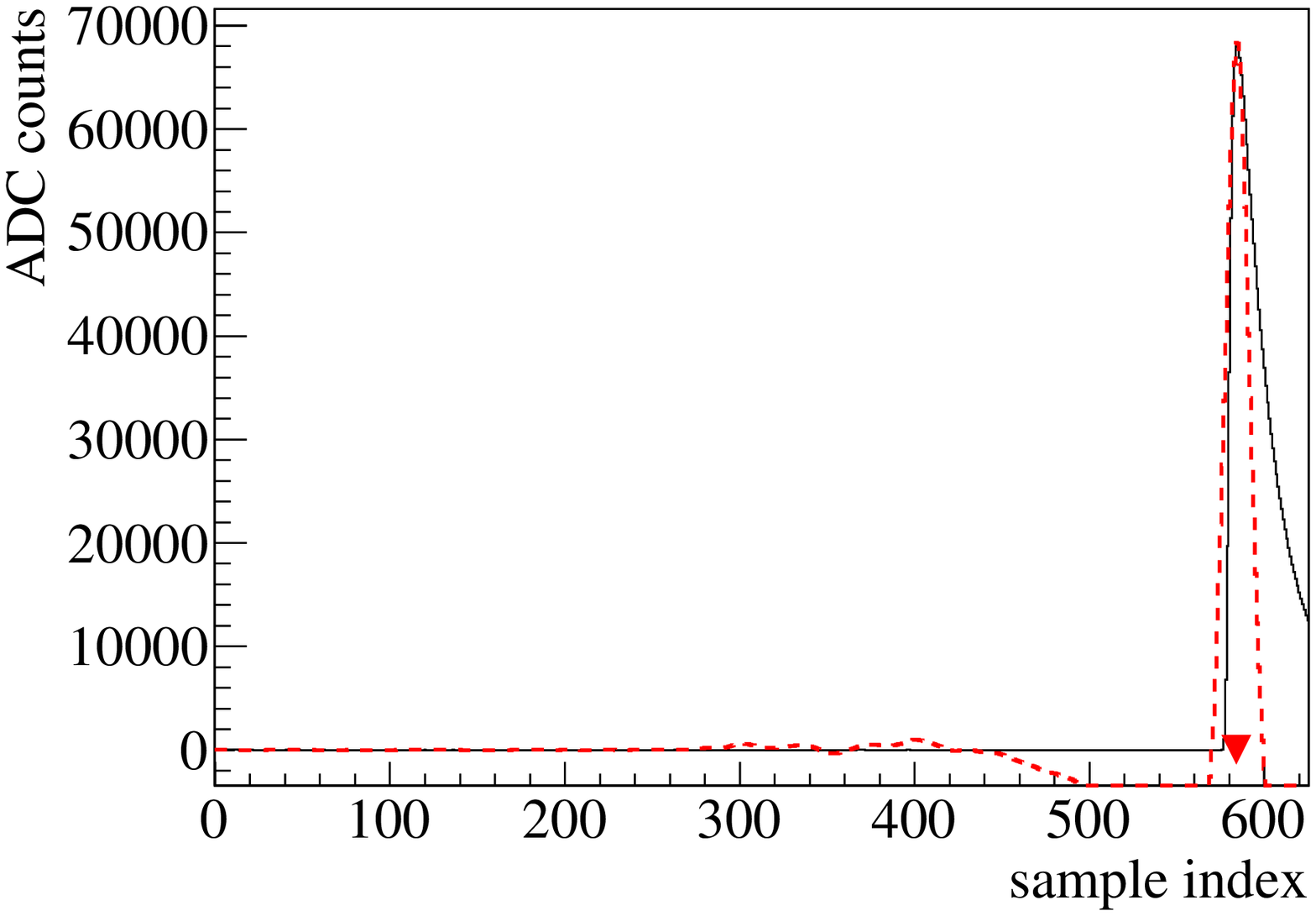}
\end{minipage}
\begin{minipage}{0.49\textwidth}
\includegraphics[clip=true,width=1\textwidth]{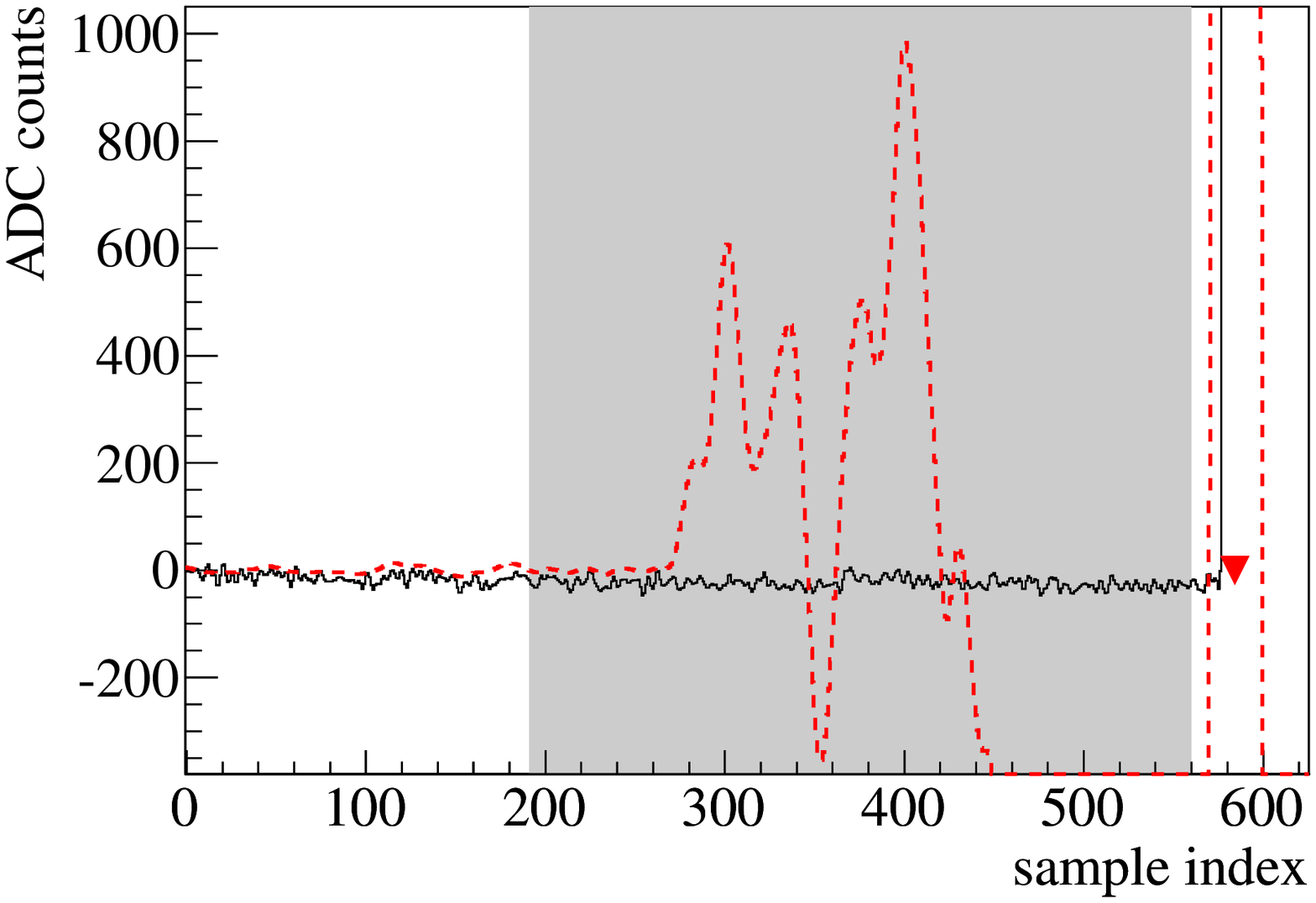}
\end{minipage}
\caption{Filtering of a high-energy signal (5407\un{keV} $\alpha$). 
When a high energy pulse occurs (left) the side lobes of the
filtered signal exceed the threshold, and fake signals could be triggered.
To avoid this, the regions corresponding to the lobes are vetoed and no
fake trigger is saved.  In the right picture: the left lobe of the pulse,
and the vetoed region (gray band).}
\label{fig:of_rinculi}
\end{figure}

Although the matched filter dramatically simplifies the implementation
of trigger algorithms, there is a drawback. When high-energy pulses are
triggered, a set of secondary pulses is seen.  This is due to the fact
that the filtered pulse has two symmetric lobes above zero, that would
be marked as signal by the trigger  (see Fig.~\ref{fig:of_rinculi}).
This problem is solved as follows.  The amplitude above zero of
the side lobes is a constant fraction $\alpha$ of the amplitude  of
the main lobe, and is estimated from the filtered average pulse.
When a signal with amplitude $A > \theta/\alpha$ is triggered, the
data samples corresponding to the expected positions of the side lobes
are vetoed, and no trigger is allowed in those regions (see Fig.~\ref{fig:of_rinculi}). 
This is an important source of inefficiency for the measurements discussed in this paper, since the \PO\ contaminations were high and the amount of dead time was dominated by the rate of the 5407\un{keV} line.
When crystals are sufficiently aged (one or two years since production), the rate is so low that the dead time is negligible.

\section{Detection efficiency} 
\begin{figure}[h]
\centering
\includegraphics[width=0.7\textwidth]{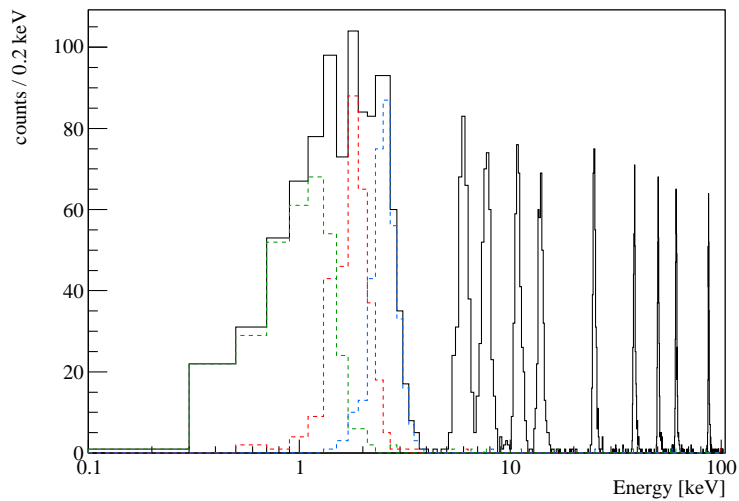}
\caption{Energy spectrum of heater pulses in the scan measurement.
  Only pulses flagged as ``heater'' by the DAQ are shown.
  Below $\sim3\un{keV}$, pulses were fired at 3 different energies.}
\label{fig:CCVR2_NPulses_ClockWork_Spectrum_ch3}
\end{figure}

To measure the detection efficiency,
an energy scan with the external heater was performed using sequences
of pulses from the ${\rm keV}$ region up to $\sim100\un{keV}$ 
(Fig.~\ref{fig:CCVR2_NPulses_ClockWork_Spectrum_ch3}).

Comparing the trigger output with the DAQ heater flag, each pulse is
considered detected if:
\begin{enumerate}
\item The trigger fired on a data sample in a time window that goes from
20 samples before the flag to 30 samples after.
\item The energy obtained from the matched filter amplitude is compatible
with the energy expected for the specific heater pulse.
\end{enumerate}
In order to verify this last condition, energy spectra
of events related to each heater amplitude set are considered. In
each spectrum there is a single peak and some outlier in which the
matched filter failed to assign the correct amplitude, due to noise
fluctuations or pile-up phenomena. Using the resolution $\sigma_f$
in Eq.~\ref{eq:of_noise_ps}, all the events not included in a $2\,{\rm
FWHM}$ energy window ($4.7\sigma_f$), centered on the mean energy of
the peak, are flagged as not detected.  The detection efficiency is then
estimated as the ratio of the events within this energy window and
the total number of heater pulses fired at that energy. 

\begin{figure}[htb]
\centering
\begin{minipage}{0.497\textwidth}
\begin{overpic}[clip=true,width=1\textwidth]{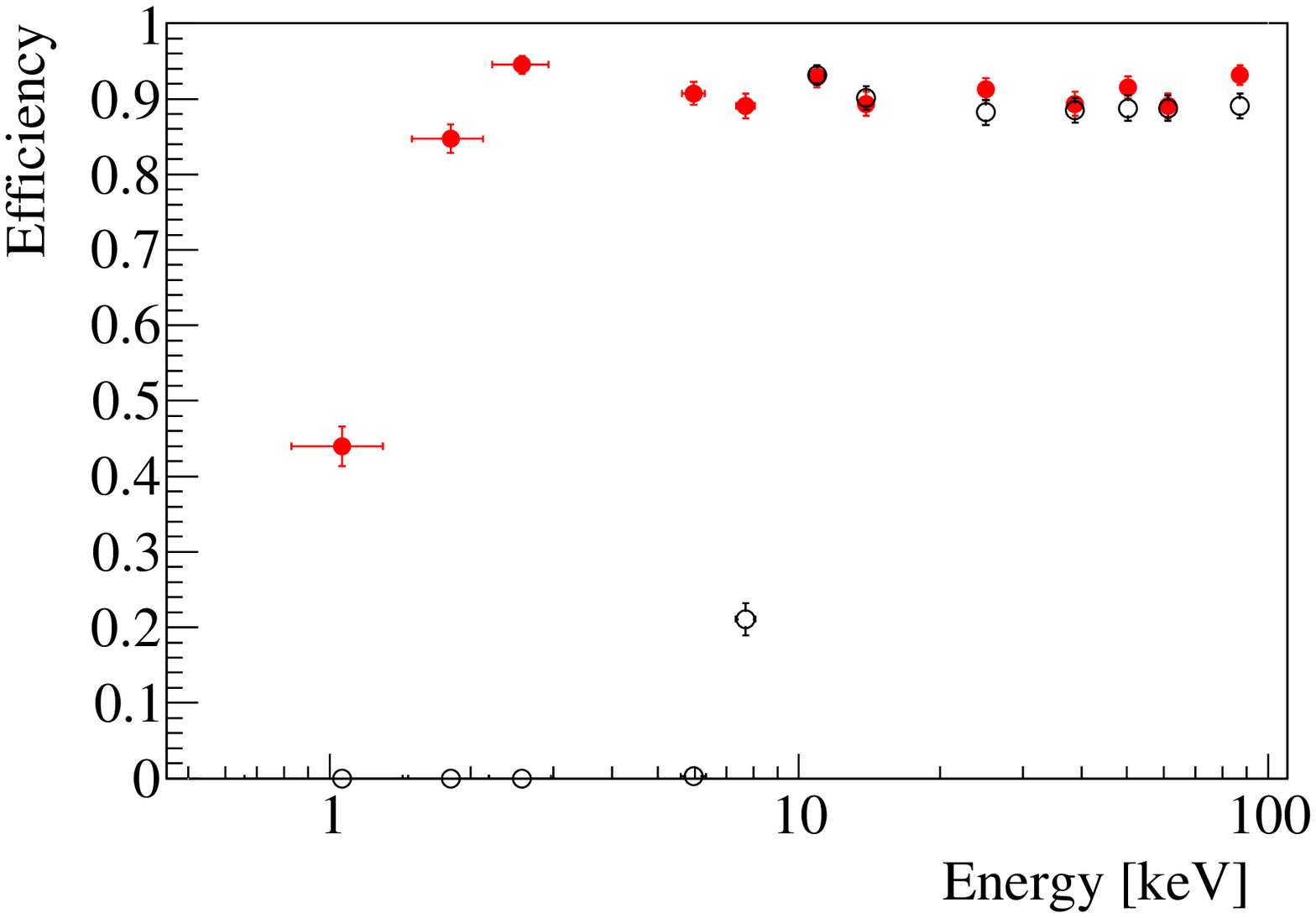}
\put(45,67){\footnotesize Bolometer 1}
\end{overpic}
\end{minipage}
\begin{minipage}{0.497\textwidth}
\begin{overpic}[clip=true,width=1\textwidth]{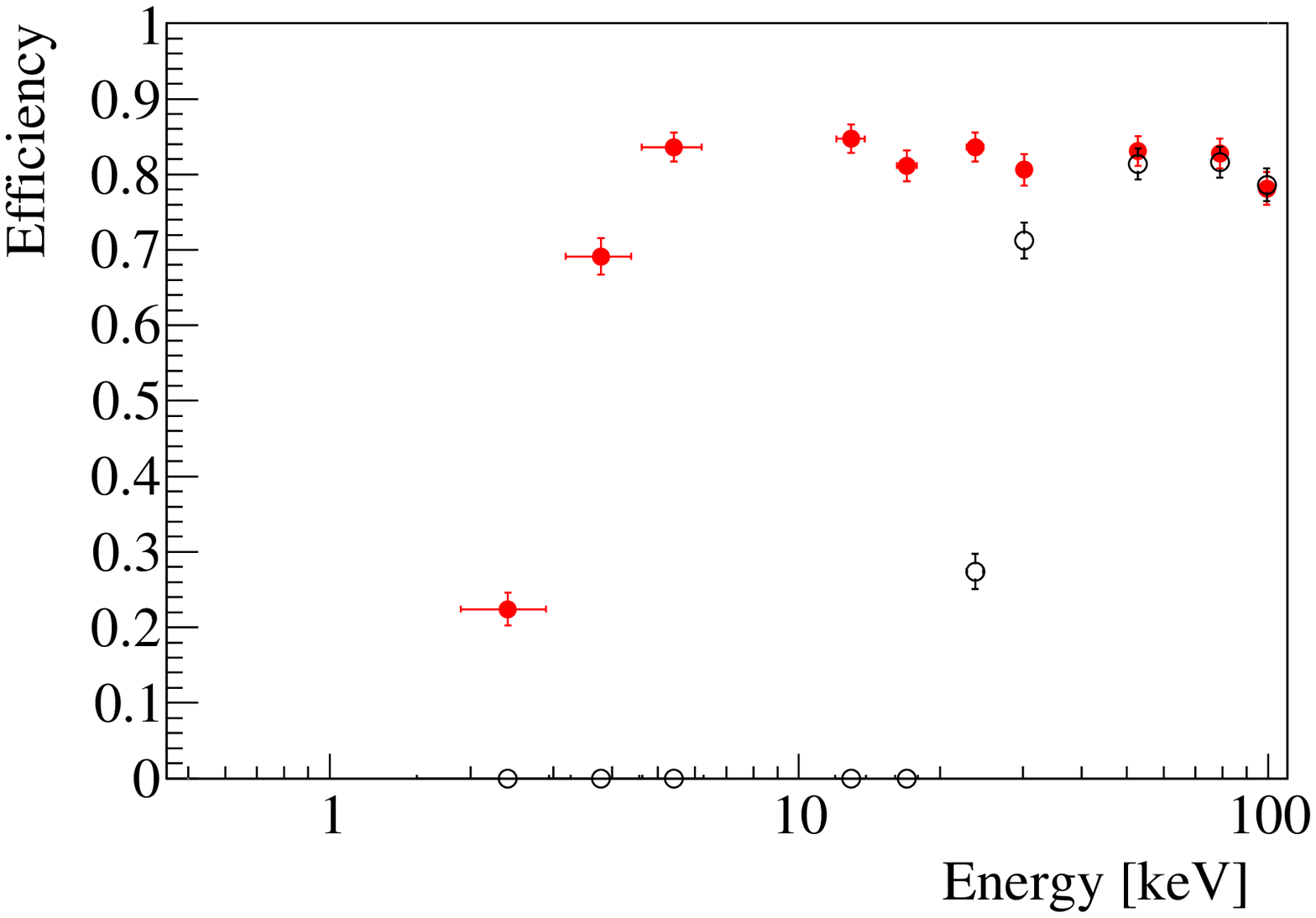}
\put(45,67){\footnotesize Bolometer 2}
\end{overpic}
\end{minipage}
\caption{Detection efficiency computed via the heater scan for the standard trigger (open black circles) and the new trigger (solid red circles).
The efficiency increases with energy, reaching a plateau.
In the first bolometer the plateau is reached at $\sim 11\un{keV}$ by the standard trigger and at $\sim 2.5\un{keV}$ by the new trigger.
In the second bolometer the plateau is reached at $30\div50 \un{keV}$ by the standard trigger and at $~\sim 5\un{keV}$ by the new trigger.
 } 
\label{fig:OTapolloheater3}
\end{figure}

The measured efficiency, compared with the one of the standard trigger, is shown
in Fig.~\ref{fig:OTapolloheater3} for the two bolometers.
The efficiency increases with energy and reaches a plateau. In the first bolometer the plateau efficiency
amounts to $90\%$ and is reached at $\sim 2.5\un{keV}$. 
This $10\%$ efficiency loss is compatible with the veto dead time due to the \PO\ line rate.
The same efficiency plateau is reached by the standard trigger at $\sim 11\un{keV}$.
In the second bolometer the improvement is even more evident: 
the $84\%$ plateau is reached at $30\div50 \un{keV}$ by the standard trigger
and at $~\sim 5\un{keV}$ by the new trigger.

\section{Pulse shape identification} 

As previously stated, the matched filter is sensitive to pulses having the
same shape as the expected signal. When a pulse with different shape occurs, its
filtered amplitude is suppressed and the filtered shape also is different
from the expectation (see Fig.~\ref{fig:ot_filt_windows}). To suppress
fake signals we implemented a pulse shape indicator.
This indicator is not used to determine the presence of a pulse, but can
be used in the data analysis to reduce the nonphysical background.

\begin{figure}[htbp]
\centering
\begin{minipage}{0.497\textwidth}
\includegraphics[clip=true,width=1\textwidth]{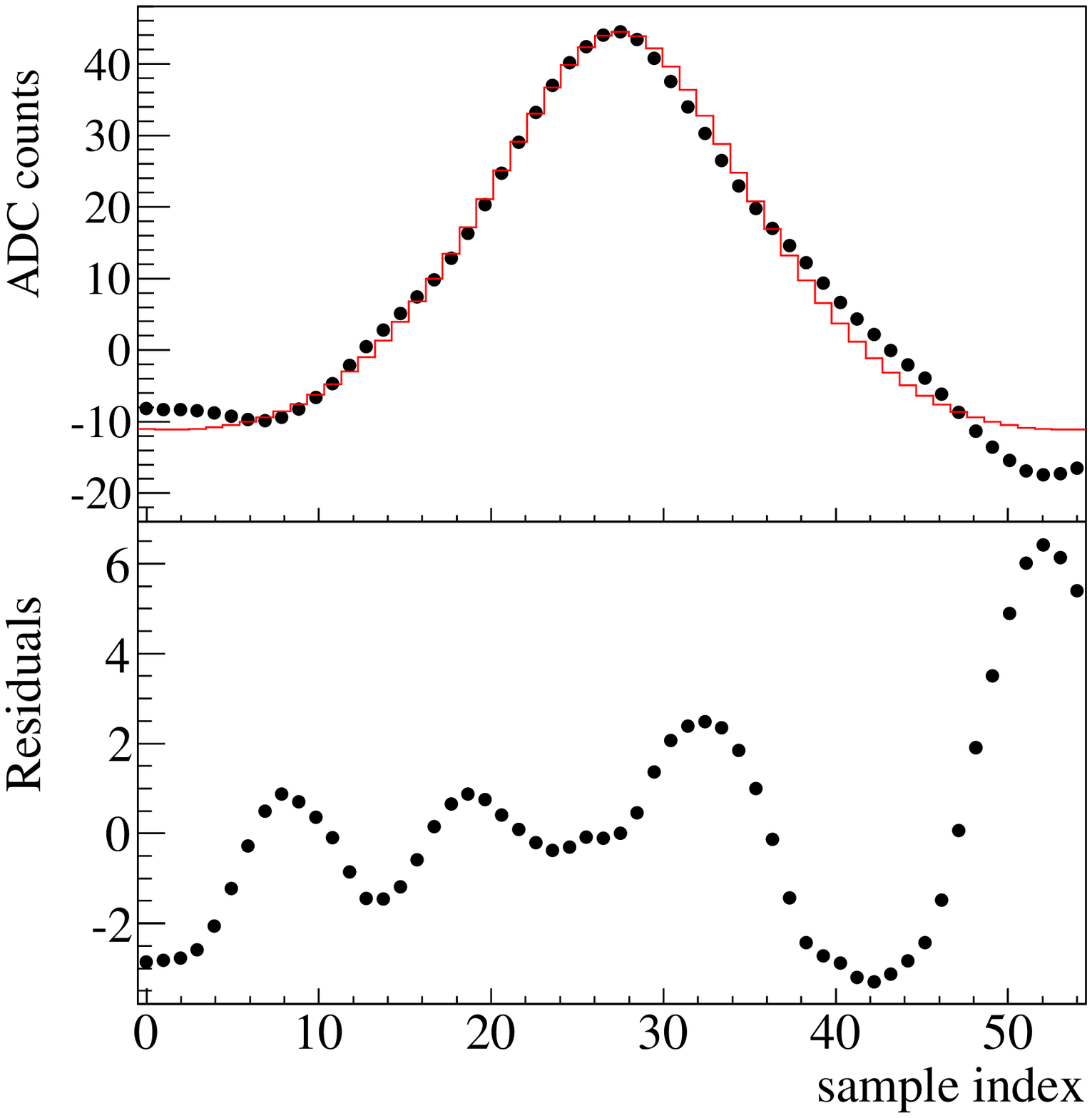}
\end{minipage}
\begin{minipage}{0.497\textwidth}
\includegraphics[clip=true,width=1\textwidth]{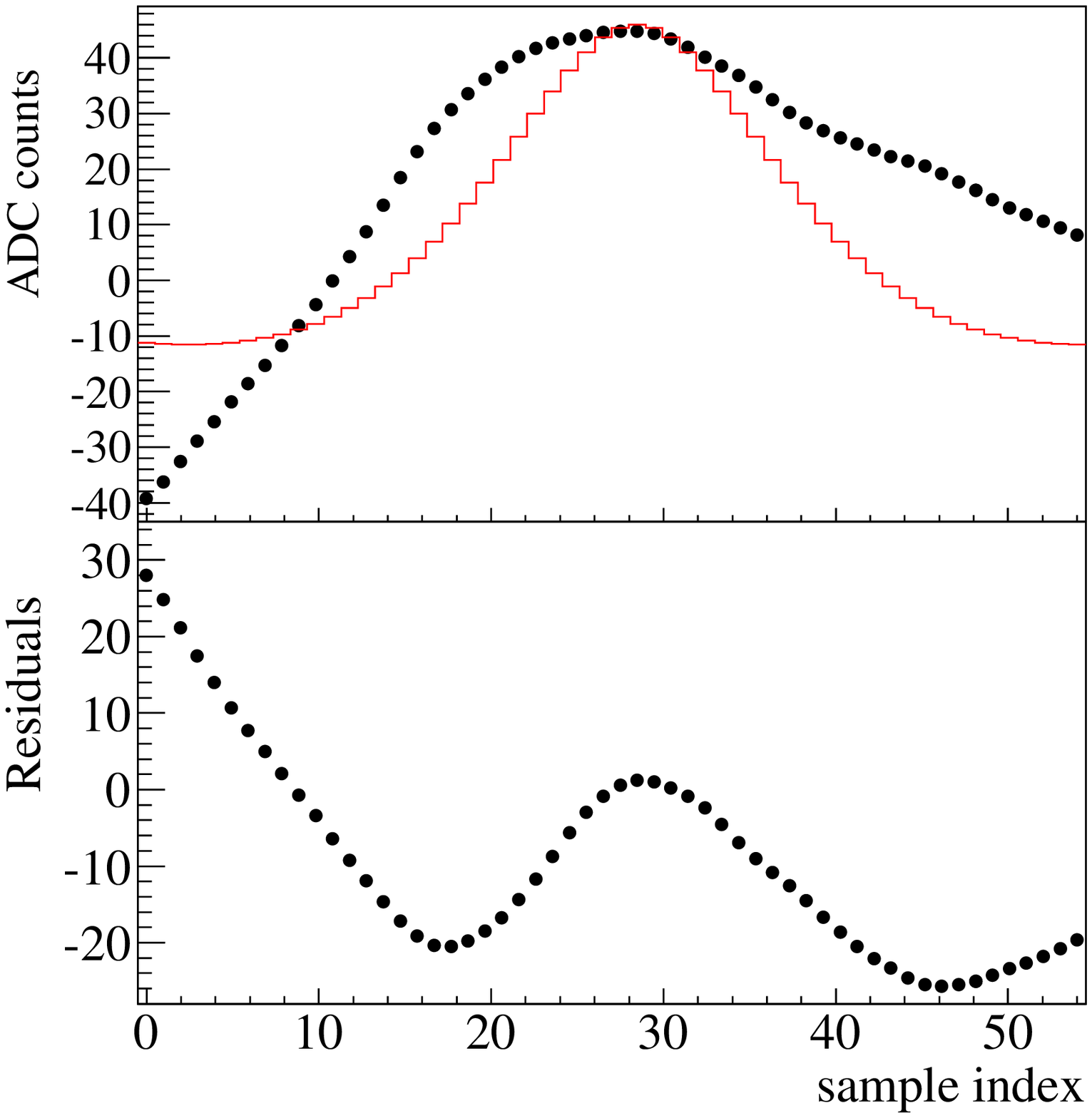}
\end{minipage}
\caption{Fits of pulses triggered on the first bolometer; points represent data and residuals, 
solid lines represent the fit functions. Left: an event in the 3-4\un{keV} region,
$SI = 0.5$. Right: another event in the 3-4\un{keV} region,
$SI = 23$.} \label{fig:of_ot_fit}
\end{figure}

The algorithm is relatively simple. A triggered pulse is fitted using a
cubic spline of the filtered average pulse, and the $\chi^2/{\rm ndf}$
of the fit is used as the shape indicator.  The cubic spline is necessary
to produce a continuous function and to fit fractional time delays.
The fitted parameters are the amplitude and the position of the pulse,
while the baseline is fixed at zero.  Since the matched filter is
itself a fit of the data samples, this fit only serves to remove the
digitization effects and to estimate the $\chi^2$. Before fitting, the
maxima of the fit function and of the triggered pulse are aligned. The
pulse position is then varied in the range $[-1,1]$ samples, which is the
maximum shift due to the digitization. The amplitude is varied in the
range $[A,A (1+\epsilon)]$, where $A$ is the amplitude of the filtered
signal and $\epsilon$ is the maximum allowed spread in amplitude due
to the digitization.  This parameter is estimated from the filtered
average pulse, computing the difference between the amplitude of the maximum
(that is equal to one) and the amplitude of the consecutive sample.
Some fits to pulses triggered on the first bolometer are shown
in Fig.~\ref{fig:of_ot_fit}.
The fit range $L$ corresponds to four FWHM of the filtered average pulse.
Finally, the shape indicator (SI) is computed as
\begin{equation}
{\rm SI} = \sum_{i=0}^{L-1} \frac{(y^f_i - f_i)^2}{\sigma_L^2 (L-2) }
\label{eq:ot_chisquare}
\end{equation}
where $y^f$ is the filtered signal, $f$ is the minimized fit function,
and $\sigma_L$ is the amount of noise expected in a window of length
$L$. The error of each point, in fact, is not the integral error of the
entire filtered window in Eq.~\ref{eq:of_noise_ps}, but it is smaller
since low frequencies are not seen in a short window:
\begin{equation}
\sigma^2_L = \sum_{k=M/L}^{M-M/L} h^2 \frac{|s(\omega_k)|^2}{N(\omega_k)}\,.
\end{equation}
It must be remarked that, even though Eq.~\ref{eq:ot_chisquare} has the
form of a $\chi^2$, it does not follow a true $\chi^2$ distribution. Although the
expected value is still 1, the variance is not $2/(L-2)$ since the
noise fluctuations are correlated at the filter output (see residuals in Fig.~\ref{fig:of_ot_fit}, left).

\begin{figure}[htbp]
\centering
\begin{minipage}{0.497\textwidth}
\begin{overpic}[clip=true,width=1\textwidth]{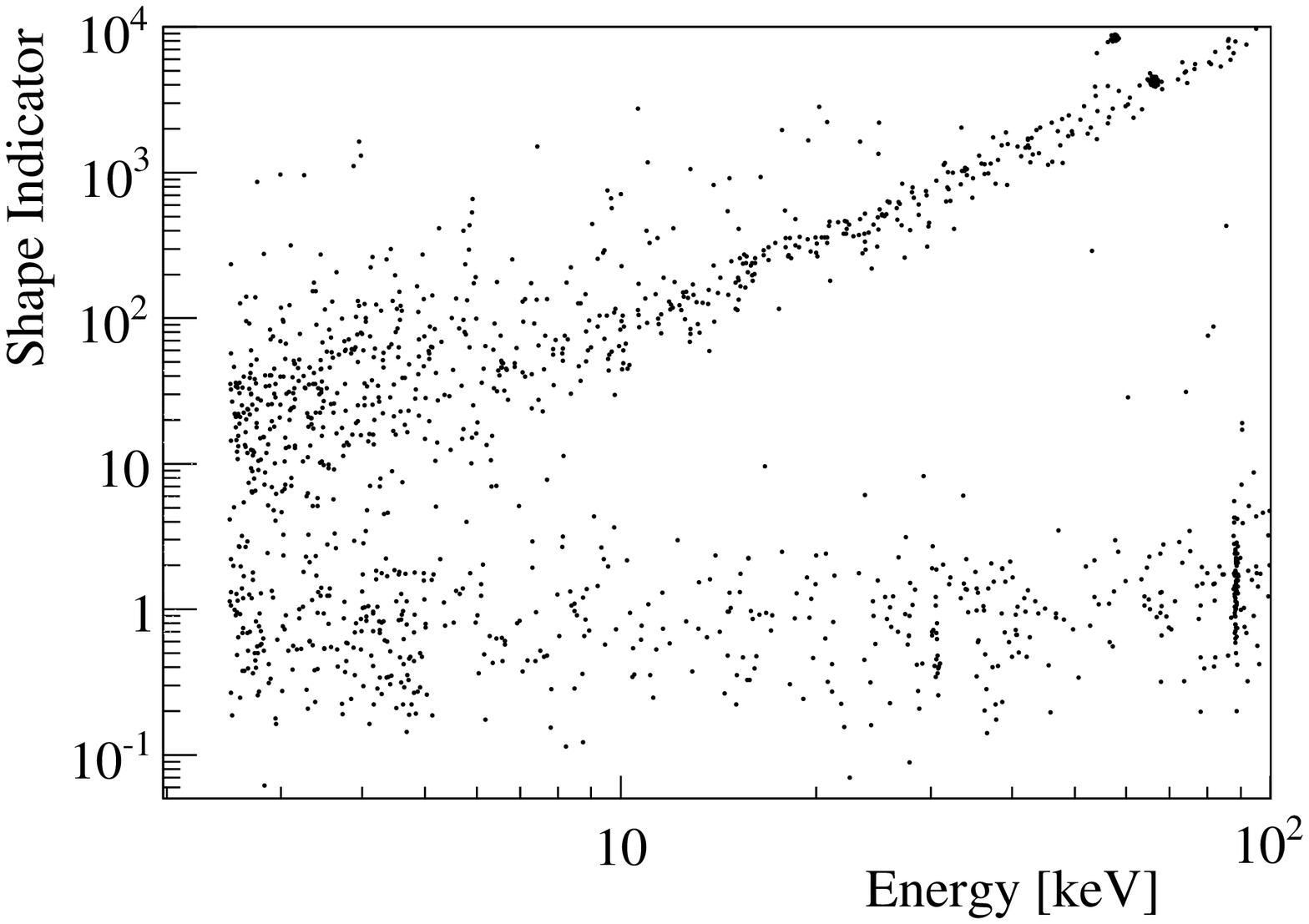}
\put(20,60){{\footnotesize Bolometer 1}}
\put(70,47){\rotatebox{25}{{\scriptsize noise}}}
\put(70,30){\rotatebox{0}{{\scriptsize signal}}}
\end{overpic}
\end{minipage}
\begin{minipage}{0.497\textwidth}
\begin{overpic}[clip=true,width=1\textwidth]{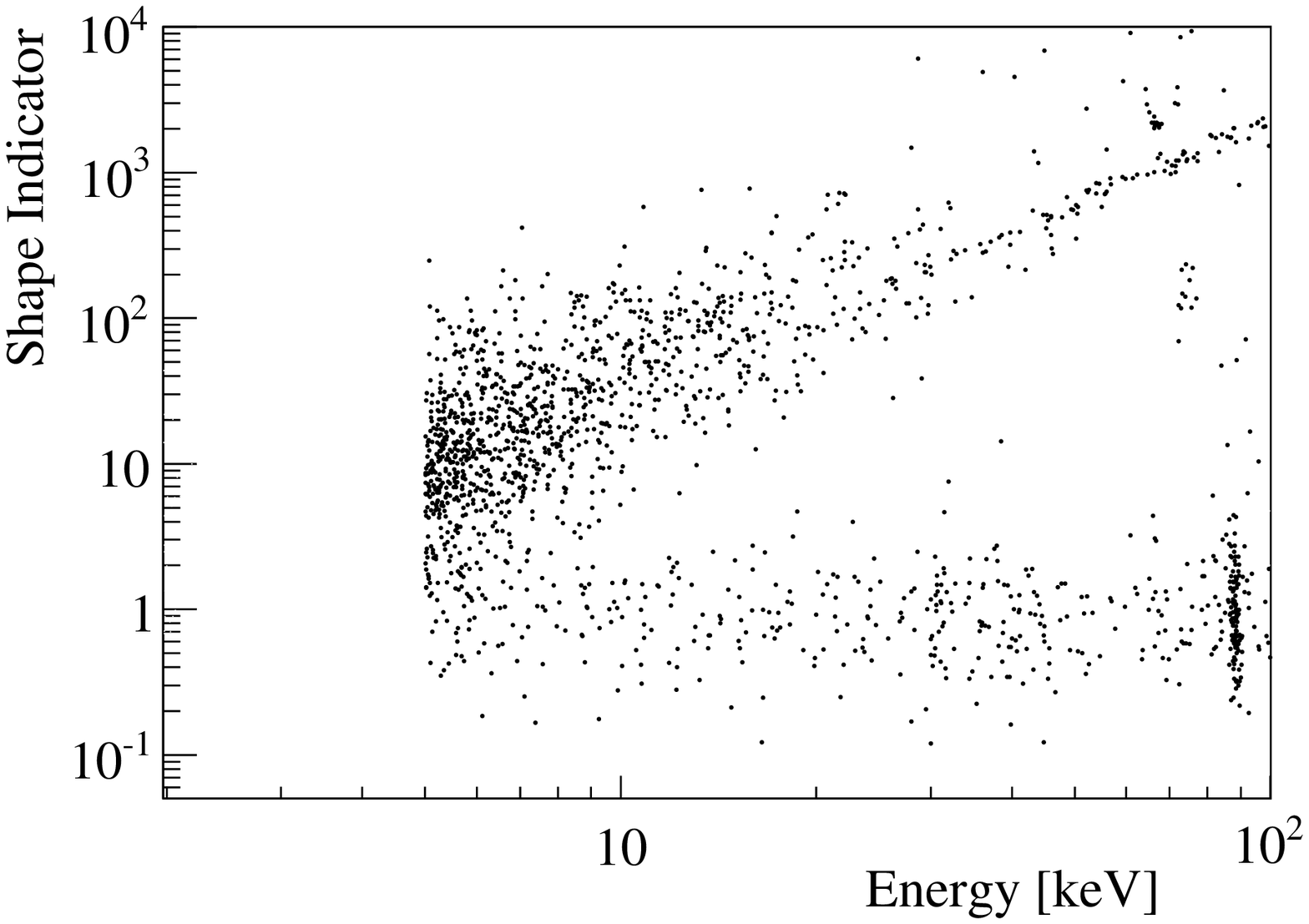}
\put(20,60){{\footnotesize Bolometer 2}}
\put(70,42){\rotatebox{25}{{\scriptsize noise}}}
\put(70,30){\rotatebox{0}{{\scriptsize signal}}}
\end{overpic}
\end{minipage}
\caption{
Shape indicator as a function of energy for the first and the second bolometer.
The band at low SI is populated by signal events, where the lines at 30\un{keV} (X-ray from Te) 
and at 88\un{keV} ($\gamma$ from $^{127m}$Te) are visible.
At higher values there are triggered mechanical vibrations and spikes.
}
\label{fig:chiotvsenergy}
\end{figure}

\begin{figure}[htbp]
\centering
\begin{minipage}{0.497\textwidth}
\begin{overpic}[clip=true,width=1\textwidth]{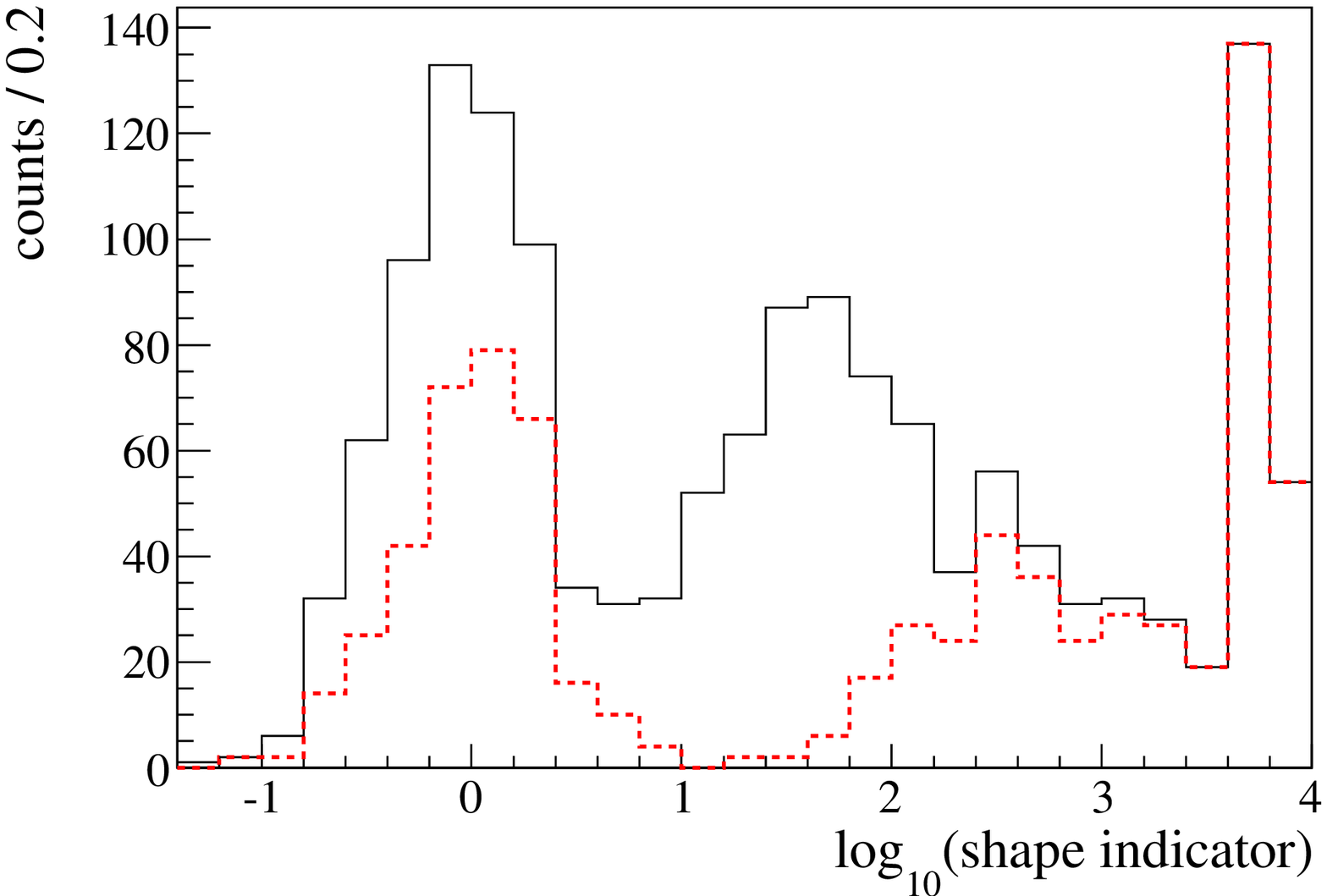}
\put(62,58){{\footnotesize Bolometer 1}}
\end{overpic}
\end{minipage}
\begin{minipage}{0.497\textwidth}
\begin{overpic}[clip=true,width=1\textwidth]{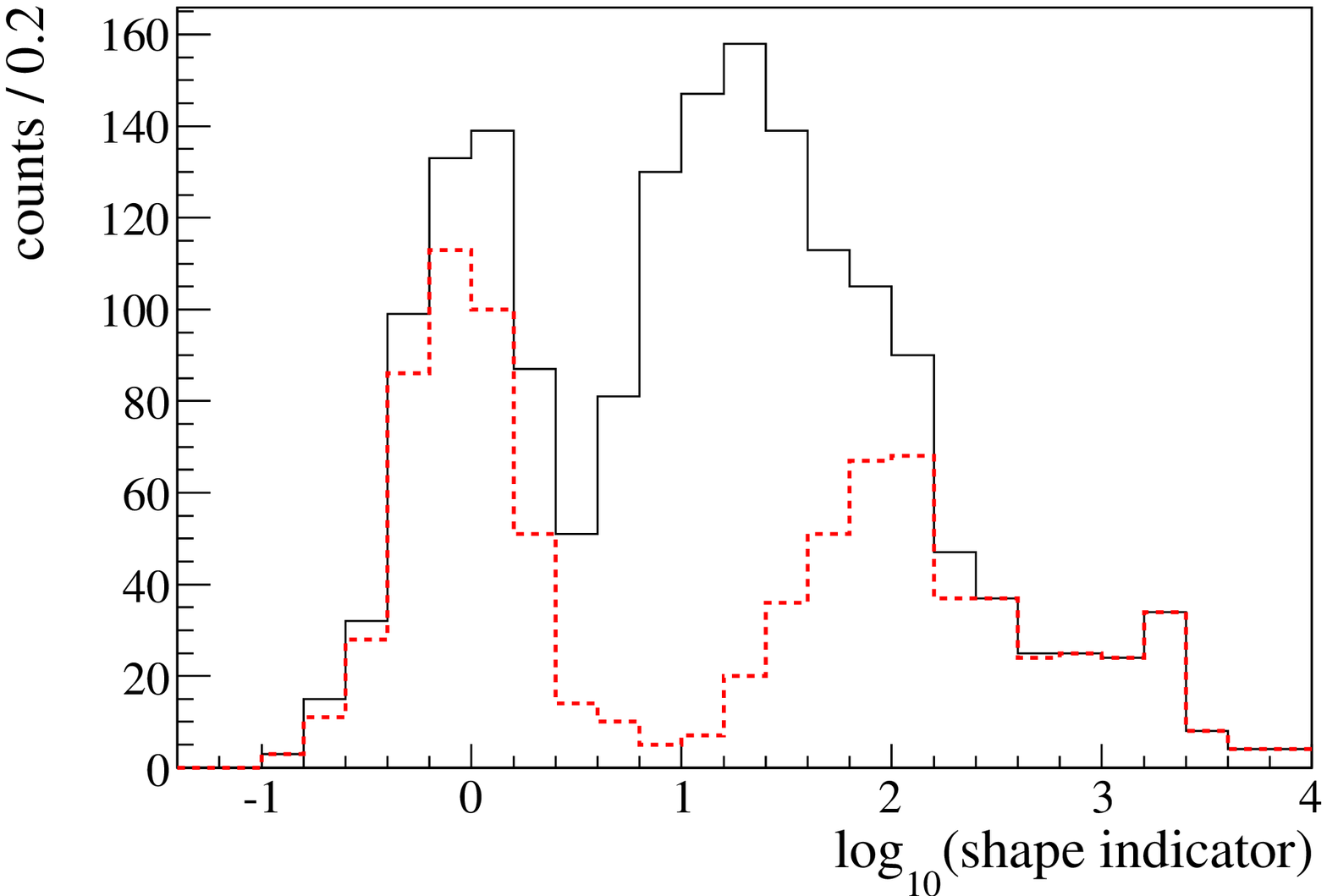}
\put(62,58){{\footnotesize Bolometer 2}}
\end{overpic}
\end{minipage}
\caption{
Shape indicator distribution for the first and the second bolometer.
The distribution of all the events in the range 2.5-100\un{keV} for the
first bolometer, and in the range 5-100\un{keV} for the second bolometer
(black solid line), shows a good separation of the signal region (peak
around $\log_{10}({\rm SI}) = 0$), which is even more evident if only the events
in the range 10-100\un{keV} are selected (red dashed line).
}
\label{fig:chiotdistro}
\end{figure}

The distribution of SI versus energy from 4.6 days of data acquired without
calibration sources is shown for both bolometers in Figs.~\ref{fig:chiotvsenergy}-\ref{fig:chiotdistro}. The signal
region is well separated, and nearly all the non-physical
pulses can be removed with a cut on this variable.
We do not discuss here the optimization of this cut, which depends on a specific analysis.

\section{Conclusions} 

We present a method to lower significantly the energy threshold of bolometric detectors, and a pulse shape parameter featuring high rejection power even at low energy.
This result is achieved by running the trigger and pulse shape algorithms on data that have been previously processed with the matched filter technique.
All the input parameters required by these algorithms can be set at optimal values automatically and no manual tuning is needed.

The application to two \TEO\ bolometers lowered their thresholds from 11 and 40\un{keV} to 2.5 and 5\un{keV}, respectively, with a detection efficiency in excess of 80\%.
The pulse shape algorithm we developed allows the rejection of nonphysical pulses, such as spikes and temperature fluctuations.
With these tools, the \TEO\ bolometers of the \Cuore\ experiment will be able to search for new physics at low energies, such as rare nuclear decays and dark matter interactions.

\acknowledgments
We are very grateful to the members of the CUORE collaboration, 
in particular to T.~Banks, C.~Brofferio, Y.G.~Kolomensky and M.~Pavan 
for their valuable suggestions, and to F.~Ferroni and M.~Pallavicini 
for supporting this work. We wish to thank S.~Frasca 
for fruitful discussions.
\bibliographystyle{JHEP} 
\bibliography{main}

\end{document}